%% file: Final_Report.tex
\documentclass[manuscript]{acmart}
\usepackage{enumitem} % For customizing list formatting
\usepackage{booktabs}    % For professional tables with better spacing
\usepackage{colortbl}    % For cell coloring
\usepackage{xcolor}      % For color definitions
\usepackage{array}    % For better table formatting
\usepackage{graphicx}
\usepackage{geometry}
\usepackage{longtable}
\usepackage{multirow}
\usepackage{array}
\usepackage{booktabs}
\setcopyright{none}
\setboolean{@ACM@nonacm}{true}
\begin{document}
\title{Human Side of Smart Contract Fuzzing: An Empirical Study}

\author{Guanming Qiao}
\email{gqiao03@ubc.ca}
\affiliation{%
  \institution{The University of British Columbia}
  \city{Vancouver}
  \country{Canada}}

\author{Partha Protim Paul}
\email{pppaul@cs.ubc.ca}
\affiliation{%
  \institution{The University of British Columbia}
  \city{Vancouver}
  \country{Canada}}
\begin{abstract}
   \textbf{Abstract} \\Smart contract (SC) fuzzing is a critical technique for detecting vulnerabilities in blockchain applications. However, its adoption remains challenging for practitioners due to fundamental differences between SCs and traditional software systems. In this study, we investigate the challenges practitioners face when adopting SC fuzzing tools by conducting an inductive content analysis of 381 GitHub issues from two widely used SC fuzzers: Echidna and Foundry. Furthermore, we conduct a user study to examine how these challenges affect different practitioner groups—SC developers and traditional software security professionals—and identify strategies practitioners use to overcome them. We systematically categorize these challenges into a taxonomy based on their nature and occurrence within the SC fuzzing workflow. Our findings reveal domain-specific ease-of-use and usefulness challenges, including technical issues with blockchain emulation, and human issues with a lack of accessible documentation and process automation. Our results provide actionable insights for tool developers and researchers, guiding future improvements in SC fuzzer tool design. 
\end{abstract}
\maketitle

\section{Introduction}
Smart contracts (SCs) have grown rapidly in recent years. By early 2023, Ethereum alone hosted over 44 million deployments, representing more than \$210 billion in market capitalization \cite{ycharts2023ethereum, qasse2023smart}. Despite its growth, many SCs lack thorough security audits, leaving them vulnerable to attacks \cite{sayeed2020smart}. A study of nearly 50,000 Ethereum contracts found widespread security issues \cite{durieux2020empirical}. To address these risks, researchers have developed various analytical approaches, including formal verification \cite{bai2018formal}, symbolic execution \cite{he2021eosafe}, and static analysis methods \cite{ma2021pluto}, however, each presents significant limitations. Formal verification requires detailed specifications in specialized languages, making it difficult to scale \cite{ma2021pluto}. Symbolic execution struggles with complex contracts due to path explosion when exploring multiple execution paths \cite{brent2020ethainter}. Static analysis often over-approximates contract behavior, leading to many false positives \cite{chen2019large}.

Fuzzing has emerged as a compelling alternative, demonstrating effectiveness in identifying vulnerabilities across traditional software platforms for many years \cite{bekrar2011finding, li2018fuzzing}. Fuzzing, when applied to SCs, dynamically explores contract behaviors, reducing false positives and uncovering hidden vulnerabilities without requiring deep prior knowledge of vulnerability patterns\cite{liang2018fuzzing}. Since 2018, numerous specialized fuzzing tools have been developed to address the unique characteristics of blockchain environments \cite{shou2023ityfuzz, grieco2020echidna, choi2021smartian, foundryrs_foundry}. 

While these tools have proliferated, little is known about the challenges practitioners face when adopting them. This study investigates the ease-of-use and usefulness barriers that affect SC fuzzing in practice. Our goal is to identify the obstacles that hinder adoption and effectiveness and to provide insights for improving tool design. In summary, this paper makes the following contributions: \begin{itemize}
    \item We develop a taxonomy of ease-of-use and usefulness challenges in SC fuzzing through an inductive analysis of GitHub issues from two widely-used SC fuzzers: Foundry\cite{foundryrs_foundry} and Echidna\cite{grieco2020echidna}.
    \item We conduct a user study with practitioners from two backgrounds—SC developers and traditional security engineers—to examine how these challenges affect their fuzzing performance.
    \item We provide empirical insights into how these challenges affect fuzzing effectiveness and document practitioner strategies for overcoming them.
    \item We discuss implications for tool design and propose recommendations to enhance the ease-of-use and adoption of SC fuzzers.
\end{itemize}
% With this proliferation of SC fuzzing tools, our research aims to illuminate the challenges practitioners encounter when learning and adopting the most widely used tools in this domain. This knowledge will help identify existing obstacles in SC fuzzing implementation and highlight promising research directions for tool improvement. Our findings derive from two primary sources. We conduct an inductive content analysis of GitHub issues reported for popular open-source fuzzing tools—Foundry \cite{foundryrs_foundry} and Echidna \cite{grieco2020echidna}—which enabled us to develop a taxonomy of user-reported usefulness and usability challenges. We then conduct a user study to examine how these challenges affect practitioners in practice, whether they manifest differently between practitioner groups, and how practitioners overcome these challenges.
\section{Related Work}
\input{related-work}

\section{Research Question}
This study aims to investigate the challenges practitioners encounter when adopting SC fuzzing tools. We organize our inquiry around the following primary research question and its supporting sub-questions:

\textbf{RQ: What challenges do practitioners face when adopting SC fuzzers?}\label{RQ:RQ}

\textbf{Sub-RQ1:} What domain-specific \textit{usefulness and ease-of-use} challenges emerge in SC fuzzing that are not present in traditional software fuzzing?\label{RQ:Sub-RQ1}

\textbf{Sub-RQ2:} How do these challenges affect the practitioners' ability to effectively fuzz SCs, and is there any difference between practitioner groups (Group A: \textit{SC developers}, Group B: \textit{Traditional software security practitioners})\label{RQ:Sub-RQ2}

\section{Methodology Design}
\subsection{Inductive Content Analysis}\label{Methodology:Inductive}
We conducted an inductive content analysis of GitHub issues from two widely used SC fuzzing tools: Echidna \cite{grieco2020echidna} and Foundry \cite{foundryrs_foundry}. Our goal was to identify domain-specific challenges in SC fuzzing not commonly found in traditional software fuzzing, thereby addressing \textbf{RQ} and \textbf{Sub-RQ1} [\ref{RQ:Sub-RQ1}]. In total, we analyzed 381 GitHub issues.

\subsubsection{Data Collection} We retrieved all open and closed issues from the Echidna and Foundry repositories with the keywords "fuzz" and "invariant", using the GitHub API. We included closed issues for two reasons. First, we observe that most usability issues are reported as questions from users seeking help. Developers would individually help onboard the user before closing the issue, without addressing the underlying ease-of-use challenges. Second, Echidna and Foundry are among the most feature-rich SC fuzzers, and offer a wide range of user-reported challenges that may inform the design of future fuzzers in academia.

\subsubsection{Human Review of GitHub Issue Summaries} 
To efficiently process the dataset, we employed three state-of-the-art LLMs to generate summaries for each issue: (i) DeepSeek R1 \cite{deepseek}, (ii) LLaMA \cite{llama}, and (iii) Mixtral Instruct \cite{mistral}. Details of our LLM pipeline and system prompt are provided in Appendix [\ref{appendix:pipeline}].
We evaluated each model by comparing its summaries of 20 randomly selected GitHub Issues against human interpretation of those issues. DeepSeek R1 produced the most accurate and comprehensive summaries and was thus selected for summarizing the remaining issues. 
% To efficiently process the large volume of GitHub issues, we leveraged three state-of-the-art large language models (LLMs) to generate summaries of each issue: i) Deepseek R1 \cite{deepseek}, ii) LLaMA \cite{llama}, and iii) Mixtral Instruct \cite{mistral}. We included our LLM-summarization pipleine and system prompt in Appendix [\ref{appendix}]. 
% After generating the summaries, we conducted a manual validation process to assess their quality and accuracy. We randomly selected 20 issues and compared the summaries generated by each model against the original issue content. Our evaluation revealed that, among three models, Deepseek R1 consistently produced the most accurate and comprehensive summaries. Consequently, we selected Deepseek R1 for summarizing the remaining issues. 
To ensure the quality of our dataset, two authors manually validated all summaries generated by Deepseek R1. This step was critical for: (1) Correcting any inaccuracies or omissions in the summaries; (2) Filtering out issues unrelated to fuzzing challenges (e.g., general tool discussions or non-fuzzing-related bugs). This manual review process ensured that our final dataset consisted exclusively of issues relevant to the challenges of SC fuzzing. We observe that LLMs did save researchers' time in this case by rapidly generating initial summaries.

% We observe that LLMs did not save researchers' time in this case, as we could not fully trust their output and had to manually go through all 381 GitHub issues.

\subsubsection{Taxonomy Development} We used card sorting\cite{spencer2009card}, a well-established method for organizing qualitative data into meaningful categories, to inductively discover challenge categories. Two researchers collaboratively analyzed the issues, which ensured that the taxonomy was grounded in multiple perspectives and minimized individual biases. For each issue, the researchers categorized it along two dimensions:
\begin{itemize}
\item Dimension A: What is the nature or type of the challenge? 
\item Dimension B: Where did the challenge occur in the SC fuzzer workflow?
\end{itemize}
We found that Dimension A reflects generalizable challenge types applicable across technical domains, while Dimension B grounds the issues to the specific context of SC fuzzing.
%while triangulating insights from our GitHub issue analysis [\ref{Methodology:Inductive}]
\subsection{Task-Oriented User Study}
We conducted a task-oriented user study to observe how the challenges affect the practitioners' ability to effectively fuzz SCs, addressing \textbf{Sub-RQ2} [\ref{RQ:Sub-RQ2}]. With participant expertise as our only independent variable, we employed a within-subjects design in which all participants completed identical tasks targeting the same vulnerabilities in the same SC. This approach allowed us to control for task difficulty across practitioner groups. 

\subsubsection{Target Contract \& Vulnerabilities}
We selected a real-world SC that met several criteria: (1) it contains bugs that were easily discoverable through fuzzing; (2) the bugs were difficult to identify through manual inspection; and (3) the contract is concise enough to be fully explored within a 90–120 minute session. The victim contract from the 2016 DAO hack met all of these criteria \cite{dao2016attack}. This contract, resembling a programmatic bank vault, allows users to deposit and withdraw funds, and contains a reentrancy vulnerability that is subtle to the human eye but detectable by modern fuzzers. We also introduced two additional synthetic vulnerabilities into the contract to provide variety and additional challenges. A detailed list of vulnerabilities is included in the Appendix [\ref{appendix:vulnerabilities}]. To avoid ambiguity and unexplained fuzzing jargon, we clarify relevant terms in the Appendix [\ref{appedix:terminology}].
\subsubsection{Recruitment \& Participants}
We recruited participants from two distinct practitioner groups:
Group A: Security researchers with expertise in traditional software fuzzing but limited blockchain experience.
Group B: Professional SC developers with strong on-chain programming experience but minimal exposure to fuzzing.
This sampling strategy enabled us to examine how differences in technical background influence users’ experiences with SC fuzzers. Participants were recruited via academic and professional networks in both the security and blockchain communities. We attached our participants' demographic information in the Appendix [\ref{tab:participant_demographics}].
% We recruited participants from two distinct practitioner groups. The first group consisted of security researchers with expertise in traditional program fuzzing but limited blockchain experience. The second group comprised professional SC developers with a deep understanding of on-chain programming, but with minimal prior exposure to fuzzing techniques. This sampling strategy enabled us to examine how different technical backgrounds influence the perception and navigation of usability challenges. Participants were recruited through professional networks and academic contacts in both the security research and blockchain development communities.
\subsubsection{Study Design} 
Each study session lasted between 90-120 minutes. Participants could extend their session if they were making progress, or end early if they believed they had found all bugs. Each session began with a briefing that introduced the study goals and obtained written consent from participants. Participants then followed a list of seven subtasks [\ref{appendix}], with access to a pre-configured development environment containing all necessary software. In order to lower the burden for participants, the researchers physically traveled to the participants' location whenever possible. Upon completion, we conducted a semi-structured interview to assess their perceptions of fuzzing, satisfaction with the tools, and their intention to adopt SC fuzzers [\ref{appendix:subtasks}]. All sessions were text-recorded with participant consent, using a think-aloud protocol to capture verbal feedback. If participants interacted with AI tools (e.g., Copilot or Claude), those interactions were also recorded for analysis. To quantitatively measure participant performance, we developed a scoring rubric that awarded points for correctly completing elements of each subtask [\ref{appendix:fuzzing_score}]. This scoring system allowed us to compare performance across groups and identify the most challenging aspects of the fuzzing process.

\section{Results}
% SC fuzzing introduces unique challenges not observed in traditional software fuzzing, primarily due to the stateful execution model of blockchains, constraints imposed by the EVM, gas usage limitations, and complex multi-contract interactions. These challenges reflect fundamental differences between SCs and conventional software systems.

\subsection{Sub-RQ1: Domain-Specific Usefulness / Usability Challenges in SC Fuzzing} The complete taxonomy is presented in Table~[\ref{tab:taxonomy}] in the appendix. Here, we focus only on the subset of challenges that are specifically relevant to SC fuzzers. Drawing from our complete taxonomy [\ref{tab:taxonomy}], we present the key domain-specific challenges as follows:
\begin{enumerate}
    \item \textit{Cross-Chain Compatibility Issues}. Even though Foundry and Echidna promise general compatibility with all derivative blockchains that implement Ethereum Virtual Machine (EVM) like Optimism, Arbitrum, Polygon, and BSC, the subtle differences between these chains' implementation could introduce cross-chain compatibility issues. Significant effort is required to handle cross-chain edge cases. As an unaddressed open issue documents, "developers will run into issues due to the fact that not all EVM compatible networks have the same semantics as L1"\cite{cross-chain}. 
    
    \item \textit{Gas Constraints}. SCs operate under strict gas limit, an EVM terminology for the fee incurred when running transactions, causing fuzzers to generate false positives or negatives if they fail to accurately simulate the gas consumption of generated transactions. This differs fundamentally from traditional fuzzing, where the number of operands fuzzed is not a concern for accuracy. An issue called for options to control the maximum transaction fee tolerable by adding `vm.callTimeout` and `vm.setGas` configurations \cite{gas-limit}. Others report inaccurate gas accounting \cite{foundry2023issue7313, foundry2024issue8789}.
    
    \item \textit{State Management And On-Chain Fuzzing Issues}. The SC fuzzers emulate a local blockchain, supporting features like chain forking, state updates and rollbacks. We observe reliability issues arising from this. Some users observe leakage of blockchain state between tests, tests running on corrupt blockchain states, and fuzzer crashes due to internal state update failures \cite{state1, state2}. Others found that inconsistencies in blockchain state storage cause test results to become unreplayable\cite{state3}. Further, on-chain fuzzing introduces additional complexity with slower RPC calls compared to traditional fuzzing where all data is locally available\cite{state4}. 

    \item \textit{Manual Yet Inflexible Fuzzer Setup Process}. Compared to traditional fuzzers, running SC fuzzers require more manual steps. Even though we did observe issues directly requesting more automation\cite{config1, foundry2023issue5466}, more issues indirectly reflect this challenge by requesting clarification for the manual setup process\cite{docum3, foundry2024issue8934}. We observed in the issue replies that developers had to individually walk the user through manual setup, even writing the fuzz harness on the user's behalf.
    \begin{quote}
        Foundry core maintainer: "What you need to do is to define a handler and mark it as targeted contract[...], so in your case pushToken is called as fuzzed selector[...]". \cite{foundry2024issue8934}
    \end{quote}
    Despite the onerous manual setup requirements, the users are rewarded with little customizability over their fuzzing campaign. Issues have reported a lack of fine-grained control over per-test timeouts, max executions per function, or logic to continue fuzzing after initial crashes\cite{foundry2024issue9393, foundry2022issue990}. For advanced users, the fuzzers also do not support adding custom logic in the core fuzzing loop, such as a custom scheduler, mutator, or path resolver \cite{foundry2024pull10190}.\label{rq1-ease-of-use-1}
    
    \item \textit{Lack of SC-Specific Documentation}. Besides a lack of automation, a few issues of users seeking help with basic setup also reflected difficulty finding tutorials and concrete code examples for the expected workflow \cite{docum1, docum2, crytic2023issue980}. We observed that both Foundry and Echinda's documentation explains the core user journey abstractly or through incomplete code snippets, making application to real-world projects difficult. \label{rq1-ease-of-use-2}\vspace{-0.1cm}
\end{enumerate}
The collection of these domain-specific usability and usefulness challenges collectively contributes to a steeper learning curve and potentially hinders the widespread adoption of fuzzing techniques within the SC development lifecycle. Addressing these specific points is crucial for making SC fuzzing more accessible and effective for a broader range of developers.

% Addressing the technical challenges reported in (1-3) requires \textbf{specialized fuzzing techniques}, including \textbf{blockchain state emulation and snapshotting, transaction cost-aware fuzzing and cross-chain compatibility updates}, which are not necessary for traditional software fuzzing.
\subsection{Sub-RQ2: How Do Challenges Affect Practitioners And Any Difference Between Groups}
In the following, we present our observations from analyzing the user studies conducted with SC developers and fuzzing practitioners. Since all participants chose to work with Foundry, our findings are limited to the challenges manifested in the Foundry tool. 

As shown in [\ref{tab:fuzzing_steps}], we observed that 5 out of 6 participants reported a lack of willingness to adopt the tool. Both groups of users faced significant challenges effectively fuzzing with Foundry, with only two participants going beyond the trivial bug \#1 within 90 to 120 minutes of focused attempt. We observed that developers faced more practical challenges running the fuzzer (Finding \#1), while fuzz practitioners reported more frustration with the lack of automation (Finding \#2). We also observed that SC developers performed better with tasks that require SC domain knowledge (Finding \#3) and that AI could significantly bridge the conceptual gap SC developers experienced running the fuzzer (Finding \#4).
\begin{table}[H] 
\centering
\renewcommand{\arraystretch}{1.2}
\setlength{\tabcolsep}{6pt}
\begin{tabular}{|p{3.2cm}|c|c|c|c|c|c|c|c|}
\hline
\rowcolor{lightgray} 
\textbf{Fuzzing Steps} & \textbf{P1} & \textbf{P2} & \textbf{P3} & \textbf{Developer Avg.} & \textbf{P4} & \textbf{P5} & \textbf{P6} & \textbf{Fuzzing Practitioner Avg.} \\
\hline
Targeted right functions & 3/3 & 3/3 & 3/3 & 100\% & 3/3 & 3/3 & 3/3 & 100\% \\
\hline
Wrote fuzz harness & 1 & 1 & 0 & 67\% & 1 & 1 & 1 & \textbf{100\%} \\
\hline
Deployed peer contracts & 0 & 0 & 0 & 0\% & 0 & 0 & 0 & 0\% \\
\hline
Initialized chain state & 1 & 1 & 0 & 67\% & 1 & 0 & 1 & 67\% \\
\hline
Wrote correct invariants & 2/3 & 2/3 & 3/3 & \textbf{89\%} & 3/3 & 1/3 & 1/3 & 55\% \\
\hline
Triggered bugs & 1/3 & 2/3 & 0/3 & 33\% & 3/3 & 1/3 & 1/3 & \textbf{55\%} \\
\hline
Interpreted output & 0/3 & 2/3 & 0/3 & 22\% & 3/3 & 1/3 & 1/3 & \textbf{55\%} \\
\hline
Satisfaction & 1/5 & 2/5 & 1/5 & 26\% & 2/5 & 1/5 & 1/5 & 26\% \\
\hline
Intention to Adopt & 0 & 0 & 1 & 33\% & 0 & 0 & 0 & 0\% \\
\hline
\end{tabular}
\caption{Fuzzing Steps Completion Rates by Participant}
\label{tab:fuzzing_steps}
\end{table}
\vspace{-.7cm}
\textbf{Finding \#1: The lack of SC-specific fuzzer documentation, tutorial, and code examples pose more challenges to those without fuzzing expertise}. For SC developers, we observed a significant gap between their mental models of fuzzing and Foundry's actual workflow. SC developers either had no prior knowledge of fuzzing, or conceptualized it as simply adding randomness to unit tests. No conceptual primer or step-by-step code example is provided in the Foundry documentation to bridge this knowledge gap. The challenge was compounded by inconsistent terminology and use of unexplained jargon. Developers particularly struggled to differentiate between 'invariant testing' and 'fuzz testing', which were presented as separate techniques despite invariant testing being essentially fuzz testing with stateful sequence support. These conceptual confusions manifested in frustration and failure to follow best practices. P1 and P3 initially created test harnesses with hardcoded API sequences instead of defining flexible property boundaries that could be effectively fuzzed. Around the 30 minutes mark, all three developers abandoned attempts to conceptually understand the documentation, instead relying on AI or themselves to write and debug the task.
\begin{quote}
      It's unclear to me where the randomness happens. Don’t you need to add randomness somewhere? The documentation doesn’t even mention the word “random”. (P1)
      Documentation doesn't really help me here. It's too much. Fuzzing is not trivial. (P3)
\end{quote}

\textbf{Finding \#2: While fuzzing practitioners are better at writing fuzz harness, they report more frustration with the lack of automation.} Fuzzing practitioners P4 and P6 recognized similarities between SC fuzzing and traditional program fuzzing paradigms, correctly identifying invariants as analogous to bug oracles and harnesses as similar to fuzz targets. But they recommended harness automation based on their experience with better tools in traditional program fuzzers.
\begin{quote}
Automating the fuzz harness generation would be a good idea. You could provide a harness that fuzzes all endpoints, and let us cut it down. That at least gives us something to start with. (P4)
I never found property-based fuzzers a good idea. There is too much manual work required for adoption. Have you used a property-based fuzzer in real life? (P5)
For me, the strength of fuzzing lies in automation – hence our domain’s ACM keyword “automated testing”. If we have to write even more code than unit tests, then what’s the point of fuzzing in the first place? (P6)
\end{quote}\label{quote:user-study-finding-2}

\textbf{Finding \#3: SC developers are better invariant writers.} SC developers crafted more effective invariants by leveraging their understanding of the Decentralized Financing (DeFi) business logic. As P1 and P3 explained:
\begin{quote}
    This is a standard ERC-20 interface... I think I know what to assert. (P1)
    I am going to look up the ECR-20 standard as my invariant reference. Any behavior violating the standard is considered buggy. (P3)
\end{quote}
All three SC developers also optimally chose to focus on testing the \texttt{deposit}, \texttt{withdraw}, and \texttt{withdrawAll} endpoints, with the experienced developer P3 identifying the crashing bug \#1 without even running the fuzzer. In comparison, fuzzing practitioners took ad-hoc approaches to API selection. P4 categorized target APIs into ``data sources'' and ``data sinks,'' correctly inferring which APIs needed to be tested in sequence from a dataflow perspective, while P5 and P6 just included all APIs as targets. 

\textbf{Finding \#4: No participant successfully implemented multi-actor fuzzing patterns following best-practice recommendations} The 'Actor Pattern' is a Foundry feature that enables the fuzzer to assume multiple contract roles within a single campaign and simulate peer-to-peer interactions\cite{foundry_invariant}. This multi-actor capability represents a novel approach even for participants with fuzzing experience. Due to insufficient documentation, none of our participants successfully comprehended or implemented this technique. Instead, all participants utilized a single fuzz harness that directly interacted with the target contract, which limited testing to one exploitation pattern at a time. While this limitation did not affect fuzzing performance in our specific study context (as all target vulnerabilities were detectable through a basic client-server interaction pattern), the failure among participants to adopt this recommended approach highlights significant usability challenges in the feature's design and documentation.

\textbf{Finding \#5: LLMs do help getting users started, but is limited in critically reasoning about fuzzer's output}. AI's generative power could provide example code tailored to the user's context and bridge the gap in documentation. We observed that less-experienced developers (P1, P2) leveraged AI to gain a better understanding of the documentation, and eventually wrote effective fuzz harnesses. While the more experienced developer P3 who did not use AI failed to run the fuzzer. AI also significantly helped fuzz practitioners getting over the Solidity language barrier.
\begin{quote}
I would rather not learn another language if I didn't have to. AI was perfect for just that. (P4) 
\end{quote}
However, we observed that AI could be less helpful with tasks that require reasoning. AI is easily misled by inaccurate or incomplete fuzzer output. \textbf{Behavior A: Foundry by default only report invariant breaches as bugs.} Even when all calls fail, the fuzzer would report "all tests passed" as long as no invariant is breached. This behavior, controlled by an undocumented "fail\_on\_revert" flag, misled 3 of 5 participants who ran the fuzzer into believing that no bug was found. Even after researchers hinted that the output may be misleading, neither the participant nor their AI assistant could identify the issue. We suspect that AI is not aware of this niche behavior due to limited discussion about it online. Only experienced fuzzing practitioners (P4 and P6) independently identified this problem by carefully reading the output and realizing that 0 function call succeeded. \textbf{Behavior B: Foundry's error messages do not distinguish between fuzz harness and invariant logic errors.} If a user-defined invariant is false by default, Foundry displayed a generic ``failed to setup environment'' message rather than clearly indicating an invariant failed. Since this generic message also appears for other setup issues, both participants who encounter this issue and their AI assistants struggled to determine whether the problem originated from their harness configuration or from the invariants themselves. 
% For example, P4 created an invariant to verify that a user's balance should be greater than zero after deposit functions execute. This approach failed because Foundry checks invariants once before the first execution, when no deposits have occurred yet, leaving P4 confused about the source of the setup failure.

AI could also be misled by its own mistakes. In P2's study, when analyzing \texttt{withdraw} function that the participant suspected was vulnerable, AI mistakenly produced three trivially true assertion statements. Later, when asked to reconsider whether the function was vulnerable, AI incorrectly concluded that the function was secure because "the fuzzer reports all tests have passed", creating a circular reasoning error where one mistake leads to another.

\section{Discussion}
\subsubsection{Data triangulation between content analysis and user study}
Both SC-specific ease-of-use challenges (\ref{rq1-ease-of-use-1} and \ref{rq1-ease-of-use-2}) identified in our GitHub Issue analysis in Sub-RQ1 are corroborated by user study findings \#1 and \#2. The user study revealed differential impacts of these issues on participants. The "Lack of SC-Specific Documentation" primarily affects developers, as they struggle to grasp fuzzing concepts. This documentation gap led most developers to develop only a limited understanding of the fuzzing process and its benefits. Among the three developers, only P3 expressed interest in further exploring SC fuzzing, largely due to prior exposure to fuzzing. P1 and P2, without prior fuzzing experience, failed to recognize value in tool adoption based on their initial interactions. Conversely, while fuzzing researchers more readily bridged the SC programming gap using AI assistance, they expressed greater frustration with the "Manual Yet Inflexible Fuzzer Setup" challenge. Having experience with more sophisticated tools in other domains, these researchers found the lengthy setup process particularly disappointing, especially given the algorithmically basic nature of the fuzzer. Our study also identified additional usability challenges that were not documented in GitHub issues. Finding \#3 highlighted the comparative advantage of SC developers in formulating effective invariants, while finding \#4 documented a significant learnability barrier with the 'Actor Pattern'—a fundamental component of the SC fuzzing workflow.
\subsubsection{Recommendations for designing usable SC fuzzers}
First, more comprehensive documentation with examples from real-world SC projects would help developers better understand and appreciate the capabilities of fuzzing. Second, increased automation is critical for shortening SC fuzzer workflows toward parity with traditional fuzzing. Tools for automatic API detection \cite{alchemy_abi_parser} and invariant generation \cite{fuzzland_invgen} could facilitate more intelligent fuzzing harness generation. Recent SC fuzzers like Ityfuzz \cite{ityfuzz} also provide built-in invariants that automatically detect a set of bugs. Finally, decoupling invariant writing from fuzz harness writing could improve productivity, leveraging SC developers' expertise in writing invariants. Developers are much better equipped to reason about a contract's expected behavior than writing fuzz harness. A specialized and highly automated fuzzer that only tests for assertion failures, combined with tools for developers to write these assertions in both stateful and stateless manner, may increase adoption rates.

\section{Conclusion}
Our research highlights key usability challenges faced by developers using smart contract fuzzers, based on GitHub issue analysis and a user study. We identified a taxonomy of barriers across both ease-of-use and usefulness dimensions, showing that existing tools often fail to meet the practical needs of developers. To improve adoption, tool developers should address domain-specific complexity, enhance automation, and prioritize better onboarding and documentation. By aligning fuzzing tools more closely with user workflows and expectations, we can make them more accessible and effective for securing smart contracts. As future work, we plan to build a prototype incorporating our recommendations and evaluate its usability and effectiveness through a follow-up user study.

\bibliographystyle{ACM-Reference-Format}
\bibliography{references}

\section{Appendix}\label{appendix}
\subsection{LLM GitHub Isuse Summarization Pipeline}\label{appendix:pipeline}
\begin{figure}[H]
    \centering
    \includegraphics[width=0.48\textwidth]{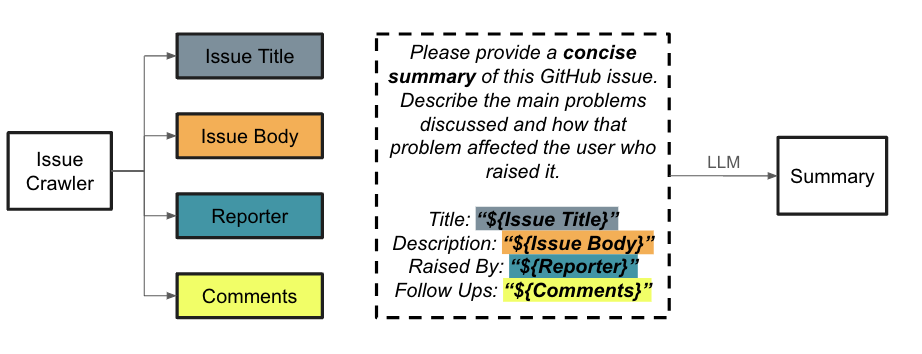}
    \caption{Pipeline for Summarizing GitHub Issues Using LLMs}
    \label{fig:summary_generation}
\end{figure}
Figure \ref{fig:summary_generation} illustrates the pipeline for summary generation. Each issue was processed by all three models to ensure robustness and to allow for comparative evaluation of the summaries.
\subsection{User Study}
\subsubsection{Subtasks Presented to User in Protocol}\label{appendix:subtasks}
\begin{enumerate}
    \item Analyze the SC to understand its purpose and identify potentially vulnerable APIs to fuzz
    \item Write a fuzz harness to target vulnerable and accessory APIs with function selectors
    \item Deploy any peer or malicious contract that interacts with the contract-under-test
    \item Set the blockchain to an initial state ready for fuzzing
    \item Develop test cases and assertions that you would like the fuzzer to check
    \item Compile and run the fuzzer to trigger the vulnerabilities
    \item Interpret trace outputs and fix the identified vulnerabilities
\end{enumerate}
\subsubsection{Fuzzing Score}\label{appendix:fuzzing_score}
\begin{enumerate}
    \item Identified the purpose of contract-under-test and chose useful function sequences to fuzz
    \item Wrote a useful fuzz harness that targets the identified functions
    \item Wrote and deployed a peer contract (normal or malicious)
    \item Initialized a correct initial chain state
    \item Wrote a correct invariant for one of the bugs
    \item Triggered the bugs
    \item Correctly interpreted the output and fixed one of the bugs
\end{enumerate}
\subsubsection{Terminology}\label{appedix:terminology}
\begin{enumerate}
    \item \textit{Target Contract.} We define target contract as the SC-under-test that contains vulnerabilities. This corresponds to the term "target program" in fuzzing literature.
    \item \textit{Target Function.} We define this as any function that contains a vulnerability, or any supporting function that must be invoked before or after the vulnerable function to successfully trigger the vulnerability.
    \item \textit{Fuzz Harness.} We define harness as the setup code that guides the fuzzer in determining which APIs to call in randomized sequences. This component is also commonly referred to as a "fuzz driver" in fuzzing literature.
    \item \textit{Invariants.} We define this as the failure conditions and assertion statements that a fuzzer evaluates after each execution. The primary objective of fuzzing is to identify execution traces that violate these pre-defined invariants. In fuzzing literature, this component is also commonly referred to as "fuzzing oracles" or "test properties."
\end{enumerate}
\subsubsection{Vulnerabilities}\label{appendix:vulnerabilities} The basic vulnerabilities of the CTF challenge are listed in Table \ref{tab:vulnerability_levels}.
\begin{table}[H] 
\centering
\renewcommand{\arraystretch}{1.3}
\setlength{\tabcolsep}{8pt}
\begin{tabular}{|c|p{3cm}|p{7cm}|}
\hline
\rowcolor{lightgray}
\textbf{Level} & \textbf{Vulnerability Type} & \textbf{Description} \\
\hline
1 & Simple crash & A basic vulnerability that triggers on any call to the function \\
\hline
2 & Off-by-one error & A subtler vulnerability causing minor loss during transactions with peer contracts \\
\hline
3 & Reentrancy attack & A complex real-world vulnerability causing severe loss during transactions with malicious peer contracts \\
\hline
\end{tabular}
\caption{SC Vulnerability Levels in CTF Challenge}
\label{tab:vulnerability_levels}
\end{table}
\subsubsection{Participants Demographics} The demographic information for the participants is presented in the table \ref{tab:participant_demographics}.
\begin{table}[H] 
\centering
\renewcommand{\arraystretch}{1.2}
\setlength{\tabcolsep}{6pt}
\begin{tabular}{|p{1.5cm}|p{2.8cm}|p{2cm}|p{2cm}|p{3.5cm}|}
\hline
\rowcolor{lightgray}
\textbf{Participant} & \textbf{Background} & \textbf{Fuzzing Exp. (Years)} & \textbf{SC Exp. (Years)} & \textbf{Demographic} \\
\hline
P1 & SC Developer & None & 2+ & Male, Asian, PhD \\
\hline
P2 & SC Developer & None & 2+ & Male, Asian, Masters \\
\hline
P3 & SC Developer & None & 2+ & Male, Asian, Bachelor \\
\hline
P4 & Researcher in Fuzzing & 2+ & None & Male, Asian, PhD \\
\hline
P5 & Researcher in Fuzzing & 1 & None & Female, Asian, PhD \\
\hline
P6 & Researcher in Fuzzing & 1 & None & Female, Asian, Masters \\
\hline
\end{tabular}
\caption{Participant Demographics and Experience}
\label{tab:participant_demographics}
\end{table}
\subsection{SC Fuzzing Workflow}
\begin{figure}[H]
    \centering
    \includegraphics[width=0.78\textwidth]{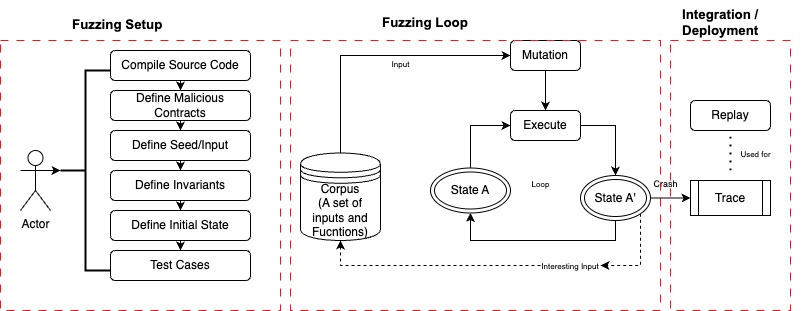}
    \caption{SC Fuzzing Workflow Overview}
    \label{fig:fuzzing-workflow}
\end{figure}

The diagram \ref{fig:fuzzing-workflow} illustrates a typical fuzzing process, particularly relevant for smart contract analysis.
        It is divided into three main phases: Fuzzing Setup, Fuzzing Loop, and Integration/Deployment.

        \textbf{Fuzzing Setup (leftmost section):} An 'Actor' initiates the setup process, which involves several sequential steps:
        \begin{itemize}
            \item \textbf{Compile Source Code:} The smart contract source code is compiled.
            \item \textbf{Define Malicious Contracts:} Potentially harmful or vulnerable contracts are defined.
            \item \textbf{Define Seed/Input:} Initial input data (seeds) are specified to start the fuzzing.
            \item \textbf{Define Invariants:} Properties or conditions that should always hold true during execution are defined.
            \item \textbf{Define Initial State:} The starting state of the system or contract is established.
            \item \textbf{Test Cases:} Specific test cases might be manually defined in addition to the fuzzing process.
        \end{itemize}
        \textbf{Fuzzing Loop (central section):} This is the iterative core of the fuzzing process. It involves:
        \begin{itemize}
            \item \textbf{Corpus (A set of inputs and Functions):} A collection of input data and function calls is maintained.
            \item \textbf{Input:} An input is selected from the corpus.
            \item \textbf{Mutation:} The selected input is mutated or modified to generate new test inputs.
            \item \textbf{Execute:} The mutated input is used to execute the smart contract, transitioning from a 'State A' to a potentially new 'State A''.
            \item \textbf{Loop:} The execution process continues, potentially leading back to 'State A' for further fuzzing.
            \item \textbf{Interesting Input:} If the execution with a mutated input leads to a new or interesting state (e.g., violating an invariant or triggering different code paths), this input might be added back to the corpus to guide further fuzzing.
        \end{itemize}

        \textbf{Integration / Deployment (rightmost section):} This phase deals with the results of the fuzzing:
        \begin{itemize}
            \item \textbf{Crash:} If the execution of a mutated input causes a crash or unexpected behavior, this is flagged.
            \item \textbf{Trace:} The execution path and state transitions leading to a crash or interesting behavior are recorded as a trace. This trace can be 'Used for' further analysis and debugging.
            \item \textbf{Replay:} The recorded trace can be replayed to reproduce the crash or the interesting behavior for investigation.
        \end{itemize}

\subsection{Distribution of 'Usefulness' and 'Ease of use' based on fuzzing workflow}
\begin{figure}[t]
    \centering
    \includegraphics[width=0.78\textwidth]{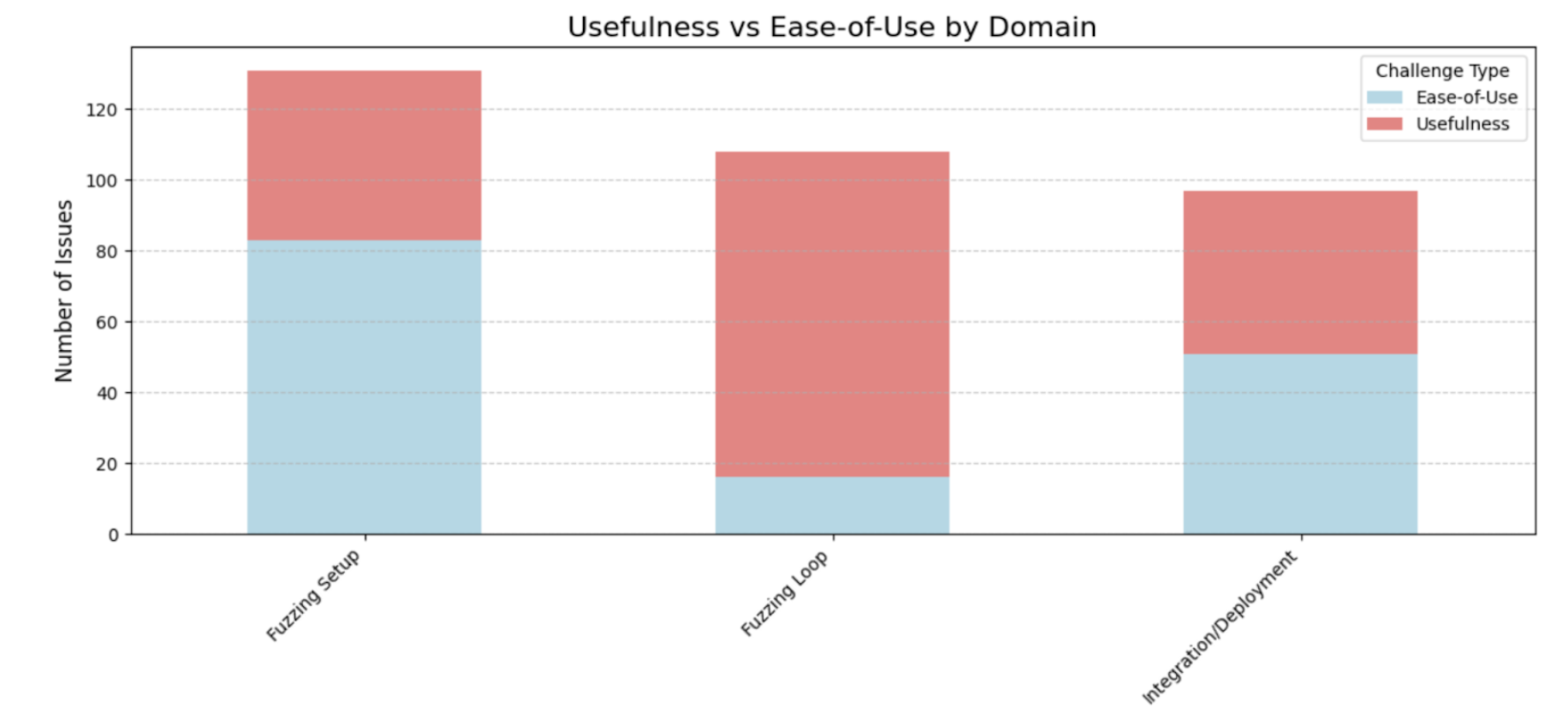}
    \caption{Distribution of Github issues based on fuzzing workflow}
    \label{fig:distribution}
\end{figure}

The stacked bar chart \ref{fig:distribution} provides a detailed breakdown of issues within three phases: Fuzzing Setup, Fuzzing Loop, and Integration/Deployment. They are categorized by "Ease-of-Use" and "Usefulness". The "Fuzzing Setup" phase exhibits the highest number of issues for both categories, indicating significant challenges in both the usability and the perceived value of this stage. In contrast, while the "Fuzzing Loop" shows fewer number of issues, "Usefulness" covers the majority. This suggests that while the process itself might be relatively easy to execute, its perceived benefit is a more prominent concern. Similarly, the "Integration/Deployment" phase presents the fewest overall issues, with "Usefulness" again outweighing "Ease-of-Use," implying that ensuring the practical value of the integration and deployment is a greater challenge than its operational simplicity.
\subsection{Challenge Taxonomy}
\input{taxonomy}
\end{document}

%% file: related-work.tex
Researchers have traditionally studied fuzzing from a technical standpoint. Many works focus on quantitative performance metrics with limited attention to human factors. Yun et al. compared embedded system fuzzers based on code coverage and crash detection rates \cite{yun2022fuzzing}. Ding and Le Goues evaluated Google OSS-Fuzz across multiple languages, including C, C++, Java, and JavaScript \cite{ding2021empirical}. Other studies examined integration challenges in CI/CD pipelines \cite{klooster2022effectiveness} and proposed evaluation frameworks for fuzzers \cite{li2021Unifuzz}. 

Although some of these studies reference "ease-of-use" or "usability," they often do so inconsistently with human-computer interaction (HCI) principles \cite{nielsenusability101}. For example, Unifuzz deemed a fuzzer "usable" if researchers could execute it successfully \cite{li2018fuzzing}. Similarly, Yun et al. briefly mentioned usability only in their conclusion \cite{yun2022fuzzing}. These studies frame utility issues from the researchers’ perspective and rarely examine real-world user experience.

More recent work has shifted toward improving fuzzer usability. Kelly et al. proposed automatic fuzz target generation to reduce developer effort \cite{kelly2019case}. Liang et al. documented fuzzing integration challenges at Huawei, including environmental constraints and training requirements \cite{liang2018fuzz}. A Shonan meeting on fuzzing research identified automation and reduced manual intervention as key goals \cite{bohme2020fuzzing}. Building on this, Nourry et al. analyzed GitHub issues for OSS-Fuzz to identify usability challenges in industrial settings \cite{nourry2023human}. Ploger et al. extended this perspective to novices, revealing barriers in learning and applying fuzzing tools \cite{ploger2021usability, ploger2023usability}. Collectively, these works highlight significant human-side challenges in fuzzing traditional programs.

However, findings from program fuzzing do not fully apply to smart contract (SC) fuzzing due to key differences in execution models and domain-specific requirements. First, SCs operate in stateful environments where previous transactions influence future behavior. In contrast, traditional fuzzers like AFL or LibFuzzer test each execution path in isolation \cite{shou2023ityfuzz}. Although stateful fuzzing exists outside blockchain (e.g., protocol fuzzing), SC fuzzers must handle complex combinations of chain state, API usage, and input values, which expands the search space considerably \cite{Natella2022, baStatefulGreyboxFuzzing}. Second, SC bugs often arise from violations of business or economic logic rather than crashes \cite{yang2025csafuzzer}. Users may need to specify invariants that reflect the intended financial behavior of the contract. Third, SCs frequently interact with other contracts in a decentralized environment. Exploits may involve multiple attacker contracts or occur when a vulnerable contract acts as a caller rather than a callee \cite{Rui2024PomaBuster, bybit2025ethcoldwalletincident}. These characteristics require users to write or integrate additional attacker and helper contracts to expose vulnerabilities.

To address these challenges, researchers have developed specialized SC fuzzers that consider the domain’s unique needs \cite{shou2023ityfuzz, grieco2020echidna, choi2021smartian, foundryrs_foundry}. Some studies have evaluated these tools based on code coverage and bug-finding capabilities. For instance, Ren et al. compared state-of-the-art SC fuzzers based on performance metrics \cite{ren2021empirical}, while Wu et al. examined how different implementation choices affect fuzzing outcomes \cite{wu2024are}. Despite these efforts, no prior work has systematically examined usability or usefulness challenges from a practitioner’s perspective in the SC fuzzing domain.

%% file: taxonomy.tex
The detailed taxonomy are presented in the table \ref{tab:taxonomy}.
\begin{longtable}{|p{0.12\textwidth}|p{0.15\textwidth}|p{0.23\textwidth}|p{0.42\textwidth}|}
\hline
\textbf{Level 1} & \textbf{Level 2} & \textbf{Level 3} & \textbf{Description} \\
\hline
\endhead

% This will appear at the bottom of each page
\hline
\multicolumn{4}{|r|}{\textit{Continued on next page...}} \\
\hline
\endfoot

% This will appear on the last page
\hline
\endlastfoot

Usefulness & Reliability & Chain State Management Issues & Failures related to managing blockchain state during testing, including forking inconsistencies, state variable manipulation problems, unexpected state leakage between tests, and challenges with storage layout resolution. \\
\hline
Usefulness & Reliability & Dependency Conflicts & Issues due to mismatches in Solidity versions, Ethereum clients, or linked libraries. Challenges related to external dependencies, including library version conflicts, failures in third-party components (like REVM), compiler compatibility issues, and dependency resolution failures in build systems. \\
\hline
Usefulness & Reliability & Cross-Chain Compatibility Problems & Difficulties when fuzzing contracts on different blockchain networks due to EVM implementation differences, varying RPC interfaces, and chain-specific behaviors that cause fork testing failures on non-Ethereum chains. \\
\hline
Usefulness & Reliability & Cheatcode and VM Manipulation Problems & Chain state setup primitives cause fuzzer crash, including timestamp manipulation incompatibilities, chain forking problems, cryptographic signing failures, and RPC call issues. \\
\hline
Usefulness & Reliability & Unstable CI/CD Integration & Fuzzers do not run smoothly in automated pipelines, making it hard to integrate into CI/CD workflows. \\
\hline
Usefulness & Reliability & Unexpected Crash During Execution & Runtime crashes, memory leaks, panics, and other execution failures that interrupt the fuzzing process and prevent complete test execution, including segmentation faults and resource exhaustion problems. \\
\hline
Usefulness & Reliability & Environment Compatibility Issues & Failures in building across different environments (Docker/NIX), and with tools behaving differently across operating systems, including segmentation faults on specific macOS versions and tests that pass on macOS but fail on Ubuntu. \\
\hline
Usefulness & Reliability & Input Generation Failures & Issues with generating valid test inputs, including type boundary violations, malformed calldata synthesis, limitations in handling constructor arguments, and failures in dictionary-based mutation strategies. \\
\hline
Usefulness & Reliability & Configuration and Feature Interaction Issues & Problems arising from conflicting configuration options, unexpected behavior when combining features, improper setting validation, and configuration parameters not being respected during execution. \\
\hline
Usefulness & Reliability & Oracle Implementation Bugs & Failures in the mechanisms that determine test outcomes, including assertion errors, false positives, incorrect error reporting, and oracle incompatibilities with state changes during test execution. \\
\hline
Usefulness & Reliability & File and Cache Management Issues & Problems with handling files and caches during fuzzing, including file locking issues in parallel execution, cache invalidation failures, directory structure errors, and artifact management challenges. \\
\hline
Usefulness & Consistency & State Management Inconsistencies & Problems with blockchain state handling during test execution, including race conditions causing false negatives, state leakage between test functions, and inconsistent state between handler and invariant contexts. \\
\hline
Usefulness & Consistency & Test Cases Not Replayable & Inability to consistently replay previously discovered bugs or execute the same fuzzing campaign with identical results, including issues with corpus replay when contracts change and non-deterministic execution across different runs. \\
\hline
Usefulness & Consistency & Non-Deterministic Behavior & Execution outcomes that vary based on the environment, including differing gas measurements between CI and local environments, platform-specific behavior variations, and inconsistent compilation outputs across systems. \\
\hline
Usefulness & Consistency & Discrepancy Within Output & Issues with how fuzzing results are reported, including discrepancies between execution traces and reported outcomes, inconsistent gas reporting, and confusing or incomplete output formats that hinder result interpretation. \\
\hline
Usefulness & Consistency & Lack of Credible Benchmark & Difficulties in establishing fair and consistent benchmarks to compare different fuzzing tools, including lack of standardized test contracts, metrics, and methodologies for performance evaluation. \\
\hline
Usefulness & Performance \& Overhead & High Gas Consumption & Fuzzing generates transactions that exceed gas limits, leading to test failures. \\
\hline
Usefulness & Performance \& Overhead & Slow Execution Due to EVM Constraints & Slow test execution due to inefficient EVM implementation, excessive transaction processing overhead, and poor utilization of available computational resources, including suboptimal thread usage and serialized execution paths. \\
\hline
Usefulness & Performance \& Overhead & On-Chain Data Access Inefficiencies & Performance degradation when fuzzing contracts that interact with on-chain data, including slow slot fetching, excessive RPC calls, poor caching mechanisms, and bandwidth limitations when retrieving blockchain state. \\
\hline
Usefulness & Performance \& Overhead & Poor Path Exploration & Fuzzer struggles to explore deep execution paths due to state-dependent logic or inefficient constraint solving. Difficulties in achieving comprehensive code coverage, including poor path exploration for complex conditions, inability to reach deep execution states, and challenges in handling state-dependent logic. \\
\hline
Usefulness & Performance \& Overhead & Input Generation and Mutation Inefficiencies & Suboptimal test case generation strategies, including redundant input creation, poor coverage of value ranges, inadequate handling of complex input types, and ineffective dictionary management. \\
\hline
Usefulness & Performance \& Overhead & Inefficient Resource Utilization & Fuzzers do not make optimal use of CPU/GPU resources during execution. Issues with effectively distributing computational resources across different test functions, handling varying test complexity, and optimizing fuzzing campaigns based on input characteristics. \\
\hline
Usefulness & Performance \& Overhead & Memory Management Issues & Excessive memory consumption during fuzzing campaigns, including unbounded growth for long test sequences, memory leaks with complex contracts, and inefficient state representation leading to out-of-memory errors. \\
\hline
Usefulness & Performance \& Overhead & Storage and Disk Utilization Problems & Challenges related to disk space usage, including excessive temporary file generation, inefficient corpus storage, and unnecessary artifact retention during fuzzing campaigns. \\
\hline
Usefulness & Performance \& Overhead & Debug and Tracing Overhead & Performance penalties when enabling debugging features, trace collection, or coverage reporting, making it difficult to simultaneously debug and maintain reasonable execution speed. \\
\hline
Usefulness & Essential Feature & Execution Control and Flow Management & Insufficient control over test execution, including lack of timeout configuration, inability to continue testing after finding failures, missing support for gas and execution time limits, and poor handling of infinite loops or hanging test cases. \\
\hline
Usefulness & Essential Feature & Limited On-Chain Oracles & Users have limited options to test on-chain events. \\
\hline
Usefulness & Essential Feature & Missing Hooks for Complex Contracts & No support for pre/post-test conditions needed for real-world DeFi contracts. \\
\hline
Usefulness & Essential Feature & Inadequate Multi-Contract Interaction & Gaps in handling complex contract interactions, including limited support for delegatecall in fallback functions, poor handling of external libraries, incompatibility with proxy patterns, and challenges with multi-contract testing scenarios. \\
\hline
Usefulness & Essential Feature & Limited Cross-Platform \& Cross-Chain Compatibility & Missing support for various development environments and platforms, including incompatibility with Windows, limited support for monorepo structures, and limited support for cross-chain testing. \\
\hline
Usefulness & Essential Feature & Blockchain State Manipulation Constraints & Limited capabilities for setting up and controlling blockchain state during testing, including inadequate ETH balance management, insufficient support for state preservation across test runs, and missing primitives for manipulating chain parameters. \\
\hline
Ease-of-Use & Quality of Life Feature & Fine-Grained Configuration Options & Difficulty configuring fuzzer parameters globally or at granular levels (per-test, per-contract), including run counts, seed formats, and test behavior options. \\
\hline
Ease-of-Use & Quality of Life Feature & Fine-Grained Input Generation and Constraint Strategies & More mechanisms for bounding inputs, specifying value ranges, handling fixtures, and incorporating values from production environments. \\
\hline
Ease-of-Use & Quality of Life Feature & Execution Control Deficiencies & Lack of fine-grained control over test execution, including handling reverts, managing specific failures, and controlling test flow. \\
\hline
Ease-of-Use & Quality of Life Feature & Preload Blockchain State & Preload contract states from mainnet/testnet snapshots to save fuzzer runtime. \\
\hline
Ease-of-Use & Error Prevention & Automate/Simplify Test Harness & Difficulty in separating test functionality from contract code, testing multi-user scenarios, and creating specialized test environments. \\
\hline
Ease-of-Use & Error Prevention & Simplify Defining Target Functions & Users struggle to specify which contract functions to fuzz. \\
\hline
Ease-of-Use & Error Prevention & Automatic Generation of Test Suite & Limited export capabilities for test results, insufficient gas reporting for fuzz tests, and challenges in generating regression tests from failures. Inaccurate or unhelpful error messages when users make mistakes writing fuzz driver/harness. \\
\hline
Ease-of-Use & Interpretability & Inadequate Non-Crash Failure Reporting & Unclear, misleading, or absent error messages when contract deployment fails or functions revert. \\
\hline
Ease-of-Use & Interpretability & Unhelpful Crash Reporting & Poor handling of arithmetic errors, unexpected opcodes, compilation issues, and other crash scenarios with insufficient debug information. \\
\hline
Ease-of-Use & System Status Visibility & Insufficient Performance Diagnostics & Limited visibility into resource utilization, workload distribution, and runtime performance bottlenecks during fuzzing. \\
\hline
Ease-of-Use & System Status Visibility & Limited Execution Tracing & Missing, incomplete, or poorly formatted execution traces, particularly for failed test cases, successful fuzzing runs. Inaccurate run counts, inadequate progress indicators, and limited status updates during fuzzing campaigns. \\
\hline
Ease-of-Use & System Status Visibility & Lack of Visualization In Bug Report & Lack of comprehensive reporting capabilities, including HTML reports, coverage visualization, progress indicators, and detailed failure information. Addresses not displayed in checksum format, and JSON output missing critical fields. \\
\hline
Ease-of-Use & Learnability \& Documentation & Hard-to-Configure Fuzzing Parameters & Confusing or unintuitive configuration formats and undocumented configuration options. For example, Foundry provides no guidance on mutation depth, seed inputs, or transaction limits, but users are expected to understand their purpose and set them on a case-by-case basis. JSON/YAML configurations are not well-documented or too rigid. \\
\hline
Ease-of-Use & Learnability \& Documentation & Undocumented Behaviors and Side Effects & Critical fuzzer behaviors not explicitly documented, including runtime contract deployment behavior, prank cheatcode effects, service contract addresses, and test case minimization logic. \\
\hline
Ease-of-Use & Learnability \& Documentation & Poor Examples for Complex Use Cases & Lack of comprehensive examples showcasing fuzzer features, particularly for complex scenarios like multi-user testing, constructor fuzzing, and edge case detection. \\
\hline
Ease-of-Use & Learnability \& Documentation & No Documented Best Practices & Lack of recommended strategies for SC fuzzing. Limited guidance on effective debugging strategies, error interpretation, and test case reproduction for fuzzing issues. Critical steps in fuzzing workflow not explicitly documented. \\
\hline
Ease-of-Use & Learnability \& Documentation & No Fuzzing Resources Provided In Documentation & Users who do not understand the basic premise of fuzzing, and fuzzing workflow have no way to learn the mental model required to work with SC fuzzers. And documentation of SC fuzzers provide no resource or links for users to educate themselves. \\
\hline
Ease-of-Use & Learnability \& Documentation & Conceptual Model Misalignment & Terminology and design choices that cause user confusion, including the dichotomy between different fuzzing approaches, unclear test patterns, and counterintuitive keywords. \\
\hline

\caption{Taxonomy of Challenges in Smart Contract Fuzzing}
\label{tab:taxonomy}
\end{longtable}

%% file: Final_Report.bbl
%%% -*-BibTeX-*-
%%% Do NOT edit. File created by BibTeX with style
%%% ACM-Reference-Format-Journals [18-Jan-2012].

\begin{thebibliography}{61}

%%% ====================================================================
%%% NOTE TO THE USER: you can override these defaults by providing
%%% customized versions of any of these macros before the \bibliography
%%% command.  Each of them MUST provide its own final punctuation,
%%% except for \shownote{} and \showURL{}.  The latter two
%%% do not use final punctuation, in order to avoid confusing it with
%%% the Web address.
%%%
%%% To suppress output of a particular field, define its macro to expand
%%% to an empty string, or better, \unskip, like this:
%%%
%%% \newcommand{\showURL}[1]{\unskip}   % LaTeX syntax
%%%
%%% \def \showURL #1{\unskip}           % plain TeX syntax
%%%
%%% ====================================================================

\ifx \showCODEN    \undefined \def \showCODEN     #1{\unskip}     \fi
\ifx \showISBNx    \undefined \def \showISBNx     #1{\unskip}     \fi
\ifx \showISBNxiii \undefined \def \showISBNxiii  #1{\unskip}     \fi
\ifx \showISSN     \undefined \def \showISSN      #1{\unskip}     \fi
\ifx \showLCCN     \undefined \def \showLCCN      #1{\unskip}     \fi
\ifx \shownote     \undefined \def \shownote      #1{#1}          \fi
\ifx \showarticletitle \undefined \def \showarticletitle #1{#1}   \fi
\ifx \showURL      \undefined \def \showURL       {\relax}        \fi
% The following commands are used for tagged output and should be
% invisible to TeX
\providecommand\bibfield[2]{#2}
\providecommand\bibinfo[2]{#2}
\providecommand\natexlab[1]{#1}
\providecommand\showeprint[2][]{arXiv:#2}

\bibitem[dao(2016)]%
        {dao2016attack}
CoinDesk \bibinfo{year}{2016}\natexlab{}.
\newblock \bibinfo{booktitle}{\emph{The {DAO} Attacked: Code Issue Leads to \$60 Million Ether Theft}}.
\newblock CoinDesk.
\newblock
\urldef\tempurl%
\url{https://www.coindesk.com/markets/2016/06/17/the-dao-attacked-code-issue-leads-to-60-million-ether-theft/}
\showURL{%
\tempurl}
\newblock
\shownote{Accessed: 2022}.


\bibitem[ych(2023)]%
        {ycharts2023ethereum}
 \bibinfo{year}{2023}\natexlab{}.
\newblock \bibinfo{title}{Ethereum Market Cap}.
\newblock \bibinfo{howpublished}{\url{https://ycharts.com/indicators/ethereum_market_cap}}.
\newblock
\newblock
\shownote{Accessed: February 2023}.


\bibitem[0xalpharush({[n.\,d.]})]%
        {docum3}
\bibfield{author}{\bibinfo{person}{0xalpharush}.} \bibinfo{year}{[n.\,d.]}\natexlab{}.
\newblock \bibinfo{title}{Confusing users by creating a dichotomy between fuzz testing and invariant testing}.
\newblock \bibinfo{howpublished}{\url{https://github.com/foundry-rs/foundry/issues/4162}}.
\newblock
\newblock
\shownote{Accessed: 2025-02-15}.


\bibitem[0xmikko({[n.\,d.]})]%
        {state2}
\bibfield{author}{\bibinfo{person}{0xmikko}.} \bibinfo{year}{[n.\,d.]}\natexlab{}.
\newblock \bibinfo{title}{Race condition in tests with false negative results}.
\newblock \bibinfo{howpublished}{\url{https://github.com/foundry-rs/foundry/issues/6450}}.
\newblock
\newblock
\shownote{Accessed: 2025-02-15}.


\bibitem[77abe77({[n.\,d.]})]%
        {state3}
\bibfield{author}{\bibinfo{person}{77abe77}.} \bibinfo{year}{[n.\,d.]}\natexlab{}.
\newblock \bibinfo{title}{State inconsistencies between handler and invariant function contexts.}
\newblock \bibinfo{howpublished}{\url{https://github.com/foundry-rs/foundry/issues/5799}}.
\newblock
\newblock
\shownote{Accessed: 2025-02-15}.


\bibitem[AI({[n.\,d.]})]%
        {deepseek}
\bibfield{author}{\bibinfo{person}{Deepseek AI}.} \bibinfo{year}{[n.\,d.]}\natexlab{}.
\newblock \bibinfo{title}{Deepseek R1}.
\newblock \bibinfo{howpublished}{\url{https://huggingface.co/deepseek-ai/DeepSeek-R1}}.
\newblock
\newblock
\shownote{Accessed: 2025-02-15}.


\bibitem[{Alchemy}(nd)]%
        {alchemy_abi_parser}
\bibfield{author}{\bibinfo{person}{{Alchemy}}.} \bibinfo{year}{n.d.}\natexlab{}.
\newblock \bibinfo{title}{ABI Parser: Decode and Encode Smart Contract Data}.
\newblock \bibinfo{howpublished}{Alchemy Website}.
\newblock
\urldef\tempurl%
\url{https://www.alchemy.com/dapps/abi-parser}
\showURL{%
\tempurl}
\newblock
\shownote{Accessed: 2025-04-09}.


\bibitem[apbendi({[n.\,d.]})]%
        {state1}
\bibfield{author}{\bibinfo{person}{apbendi}.} \bibinfo{year}{[n.\,d.]}\natexlab{}.
\newblock \bibinfo{title}{Leak into fuzz tests}.
\newblock \bibinfo{howpublished}{\url{https://github.com/foundry-rs/foundry/issues/7462}}.
\newblock
\newblock
\shownote{Accessed: 2025-02-15}.


\bibitem[Azvect({[n.\,d.]})]%
        {gas-limit}
\bibfield{author}{\bibinfo{person}{Azvect}.} \bibinfo{year}{[n.\,d.]}\natexlab{}.
\newblock \bibinfo{title}{Improved gas control}.
\newblock \bibinfo{howpublished}{\url{https://github.com/foundry-rs/foundry/issues/7892}}.
\newblock
\newblock
\shownote{Accessed: 2025-02-15}.


\bibitem[Ba et~al\mbox{.}(2022)]%
        {baStatefulGreyboxFuzzing}
\bibfield{author}{\bibinfo{person}{Jinsheng Ba}, \bibinfo{person}{Marcel B{\"o}hme}, \bibinfo{person}{Zahra Mirzamomen}, {and} \bibinfo{person}{Abhik Roychoudhury}.} \bibinfo{year}{2022}\natexlab{}.
\newblock \showarticletitle{Stateful Greybox Fuzzing}. In \bibinfo{booktitle}{\emph{31st USENIX Security Symposium (USENIX Security 22)}}. \bibinfo{publisher}{USENIX Association}, \bibinfo{address}{Boston, MA}, \bibinfo{pages}{3255--3272}.
\newblock
\showISBNx{978-1-939133-31-1}
\urldef\tempurl%
\url{https://www.usenix.org/conference/usenixsecurity22/presentation/ba}
\showURL{%
\tempurl}


\bibitem[Bai et~al\mbox{.}(2018)]%
        {bai2018formal}
\bibfield{author}{\bibinfo{person}{Xiaomin Bai}, \bibinfo{person}{Zijing Cheng}, \bibinfo{person}{Zhangbo Duan}, {and} \bibinfo{person}{Kai Hu}.} \bibinfo{year}{2018}\natexlab{}.
\newblock \showarticletitle{Formal Modeling and Verification of Smart Contracts}. In \bibinfo{booktitle}{\emph{Proceedings of the 2018 7th International Conference on Software and Computer Applications}} (Kuantan, Malaysia) \emph{(\bibinfo{series}{ICSCA '18})}. \bibinfo{publisher}{ACM}, \bibinfo{pages}{322--326}.
\newblock
\href{https://doi.org/10.1145/3185089.3185138}{doi:\nolinkurl{10.1145/3185089.3185138}}


\bibitem[Bekrar et~al\mbox{.}(2011)]%
        {bekrar2011finding}
\bibfield{author}{\bibinfo{person}{Sofia Bekrar}, \bibinfo{person}{Chaouki Bekrar}, \bibinfo{person}{Roland Groz}, {and} \bibinfo{person}{Laurent Mounier}.} \bibinfo{year}{2011}\natexlab{}.
\newblock \showarticletitle{Finding software vulnerabilities by smart fuzzing}. In \bibinfo{booktitle}{\emph{2011 Fourth IEEE International Conference on Software Testing, Verification and Validation}}.
\newblock


\bibitem[bluele({[n.\,d.]})]%
        {state4}
\bibfield{author}{\bibinfo{person}{bluele}.} \bibinfo{year}{[n.\,d.]}\natexlab{}.
\newblock \bibinfo{title}{bug(cheatcodes): calling eth\_getTransactionByHash with rpc cheatcode failed.}
\newblock \bibinfo{howpublished}{\url{https://github.com/foundry-rs/foundry/issues/7858}}.
\newblock
\newblock
\shownote{Accessed: 2025-02-15}.


\bibitem[Brent et~al\mbox{.}(2020)]%
        {brent2020ethainter}
\bibfield{author}{\bibinfo{person}{Lexi Brent}, \bibinfo{person}{Neville Grech}, \bibinfo{person}{Sifis Lagouvardos}, \bibinfo{person}{Bernhard Scholz}, {and} \bibinfo{person}{Yannis Smaragdakis}.} \bibinfo{year}{2020}\natexlab{}.
\newblock \showarticletitle{Ethainter: a smart contract security analyzer for composite vulnerabilities}.
\newblock \bibinfo{journal}{\emph{PLDI}} (\bibinfo{year}{2020}).
\newblock


\bibitem[Bybit(nd)]%
        {bybit2025ethcoldwalletincident}
\bibfield{author}{\bibinfo{person}{Bybit}.} \bibinfo{year}{n.d.}\natexlab{}.
\newblock \bibinfo{booktitle}{\emph{Incident Update - ETH Cold Wallet Incident}}.
\newblock Bybit Announcements.
\newblock
\urldef\tempurl%
\url{https://announcements.bybit.com/article/incident-update---eth-cold-wallet-incident-blt292c0454d26e9140/}
\showURL{%
\tempurl}


\bibitem[Böhme and Falk(2020)]%
        {bohme2020fuzzing}
\bibfield{author}{\bibinfo{person}{Marcel Böhme} {and} \bibinfo{person}{Brandon Falk}.} \bibinfo{year}{2020}\natexlab{}.
\newblock \showarticletitle{Fuzzing: On the exponential cost of vulnerability discovery}. In \bibinfo{booktitle}{\emph{Proceedings of the 28th ACM Joint Meeting on European Software Engineering Conference and Symposium on the Foundations of Software Engineering (FSE'20)}}. \bibinfo{pages}{713--724}.
\newblock


\bibitem[Chen et~al\mbox{.}(2019)]%
        {chen2019large}
\bibfield{author}{\bibinfo{person}{Ting Chen}, \bibinfo{person}{Zihao Li}, \bibinfo{person}{Yufei Zhang}, \bibinfo{person}{Xiapu Luo}, \bibinfo{person}{Ting Wang}, \bibinfo{person}{Teng Hu}, \bibinfo{person}{Xiuzhuo Xiao}, \bibinfo{person}{Dong Wang}, \bibinfo{person}{Jin Huang}, {and} \bibinfo{person}{Xiaosong Zhang}.} \bibinfo{year}{2019}\natexlab{}.
\newblock \showarticletitle{A large-scale empirical study on control flow identification of smart contracts}.
\newblock \bibinfo{journal}{\emph{ESEM}} (\bibinfo{year}{2019}).
\newblock


\bibitem[Choi et~al\mbox{.}(2021)]%
        {choi2021smartian}
\bibfield{author}{\bibinfo{person}{Jaeseung Choi}, \bibinfo{person}{Doyeon Kim}, \bibinfo{person}{Soomin Kim}, \bibinfo{person}{Gustavo Grieco}, \bibinfo{person}{Alex Groce}, {and} \bibinfo{person}{Sang~Kil Cha}.} \bibinfo{year}{2021}\natexlab{}.
\newblock \showarticletitle{Smartian: Enhancing smart contract fuzzing with static and dynamic data-flow analyses}. In \bibinfo{booktitle}{\emph{2021 36th IEEE/ACM International Conference on Automated Software Engineering (ASE)}}. IEEE, \bibinfo{pages}{227--239}.
\newblock


\bibitem[Crytic(2023)]%
        {crytic2023issue980}
\bibfield{author}{\bibinfo{person}{Crytic}.} \bibinfo{year}{2023}\natexlab{}.
\newblock \bibinfo{title}{Invariant Tests Can Be Skipped if Failed During Inputs Generation}.
\newblock \bibinfo{howpublished}{\url{https://github.com/crytic/echidna/issues/980}}.
\newblock
\newblock
\shownote{Issue \#980, Accessed: April 5, 2025}.


\bibitem[Ding and Le~Goues(2021)]%
        {ding2021empirical}
\bibfield{author}{\bibinfo{person}{Zhen~Yu Ding} {and} \bibinfo{person}{Claire Le~Goues}.} \bibinfo{year}{2021}\natexlab{}.
\newblock \bibinfo{title}{An Empirical Study of OSS-Fuzz Bugs}.
\newblock
\showeprint[arxiv]{2103.11518}~[cs.SE]


\bibitem[Durieux et~al\mbox{.}(2020)]%
        {durieux2020empirical}
\bibfield{author}{\bibinfo{person}{Thomas Durieux}, \bibinfo{person}{Jo\~{a}o~F Ferreira}, \bibinfo{person}{Rui Abreu}, {and} \bibinfo{person}{Pedro Cruz}.} \bibinfo{year}{2020}\natexlab{}.
\newblock \showarticletitle{Empirical Review of Automated Analysis Tools on 47,587 {Ethereum} Smart Contracts}. In \bibinfo{booktitle}{\emph{Proceedings of the ACM/IEEE 42nd International Conference on Software Engineering}} (Seoul, South Korea). \bibinfo{publisher}{ACM}, \bibinfo{pages}{530--541}.
\newblock
\href{https://doi.org/10.1145/3377811.3380364}{doi:\nolinkurl{10.1145/3377811.3380364}}


\bibitem[{Foundry}(nd)]%
        {foundry_invariant}
\bibfield{author}{\bibinfo{person}{{Foundry}}.} \bibinfo{year}{n.d.}\natexlab{}.
\newblock \bibinfo{title}{Invariant Testing - Actor Management}.
\newblock \bibinfo{howpublished}{Foundry Book}.
\newblock
\urldef\tempurl%
\url{https://book.getfoundry.sh/forge/invariant-testing#actor-management}
\showURL{%
\tempurl}
\newblock
\shownote{Accessed: 2025-04-09}.


\bibitem[Foundry-rs(2022)]%
        {foundry2022issue990}
\bibfield{author}{\bibinfo{person}{Foundry-rs}.} \bibinfo{year}{2022}\natexlab{}.
\newblock \bibinfo{title}{Fuzzer Should Respect cheatcode vm.assume}.
\newblock \bibinfo{howpublished}{\url{https://github.com/foundry-rs/foundry/issues/990}}.
\newblock
\newblock
\shownote{Issue \#990, Accessed: April 5, 2025}.


\bibitem[Foundry-rs(2023a)]%
        {foundry2023issue7313}
\bibfield{author}{\bibinfo{person}{Foundry-rs}.} \bibinfo{year}{2023}\natexlab{a}.
\newblock \bibinfo{title}{Forge Static Analysis Tool to Detect Fund Vulnerable Functions}.
\newblock \bibinfo{howpublished}{\url{https://github.com/foundry-rs/foundry/issues/7313}}.
\newblock
\newblock
\shownote{Issue \#7313, Accessed: April 5, 2025}.


\bibitem[Foundry-rs(2023b)]%
        {foundry2023issue5466}
\bibfield{author}{\bibinfo{person}{Foundry-rs}.} \bibinfo{year}{2023}\natexlab{b}.
\newblock \bibinfo{title}{Invariant Tests Stopping When Invariant Fails}.
\newblock \bibinfo{howpublished}{\url{https://github.com/foundry-rs/foundry/issues/5466}}.
\newblock
\newblock
\shownote{Issue \#5466, Accessed: April 5, 2025}.


\bibitem[Foundry-rs(2024a)]%
        {foundry2024issue9393}
\bibfield{author}{\bibinfo{person}{Foundry-rs}.} \bibinfo{year}{2024}\natexlab{a}.
\newblock \bibinfo{title}{Fuzz Error Assertion with Message Fails Test But \`forge\_snapshot\` Shows Successful Test}.
\newblock \bibinfo{howpublished}{\url{https://github.com/foundry-rs/foundry/issues/9393}}.
\newblock
\newblock
\shownote{Issue \#9393, Accessed: April 5, 2025}.


\bibitem[Foundry-rs(2024b)]%
        {foundry2024issue8934}
\bibfield{author}{\bibinfo{person}{Foundry-rs}.} \bibinfo{year}{2024}\natexlab{b}.
\newblock \bibinfo{title}{Fuzzer Stopping the Fuzzing After the invariant is broken}.
\newblock \bibinfo{howpublished}{\url{https://github.com/foundry-rs/foundry/issues/8934}}.
\newblock
\newblock
\shownote{Issue \#8934, Accessed: April 5, 2025}.


\bibitem[Foundry-rs(2024c)]%
        {foundry2024pull10190}
\bibfield{author}{\bibinfo{person}{Foundry-rs}.} \bibinfo{year}{2024}\natexlab{c}.
\newblock \bibinfo{title}{Fuzzer: Store Stats per Invariant so Failed Invariants Don't Terminate Fuzzing}.
\newblock \bibinfo{howpublished}{\url{https://github.com/foundry-rs/foundry/pull/10190}}.
\newblock
\newblock
\shownote{Pull Request \#10190, Accessed: April 5, 2025}.


\bibitem[Foundry-rs(2024d)]%
        {foundry2024issue8789}
\bibfield{author}{\bibinfo{person}{Foundry-rs}.} \bibinfo{year}{2024}\natexlab{d}.
\newblock \bibinfo{title}{Fuzzer Unable to Detect a Bug Due to Invariant With Trivial Assertion}.
\newblock \bibinfo{howpublished}{\url{https://github.com/foundry-rs/foundry/issues/8789}}.
\newblock
\newblock
\shownote{Issue \#8789, Accessed: April 5, 2025}.


\bibitem[Foundry-rs(nd)]%
        {foundryrs_foundry}
\bibfield{author}{\bibinfo{person}{Foundry-rs}.} \bibinfo{year}{n.d.}\natexlab{}.
\newblock \bibinfo{booktitle}{\emph{Foundry}}.
\newblock GitHub.
\newblock
\urldef\tempurl%
\url{https://github.com/foundry-rs/foundry}
\showURL{%
\tempurl}


\bibitem[{Fuzzland}(nd)]%
        {fuzzland_invgen}
\bibfield{author}{\bibinfo{person}{{Fuzzland}}.} \bibinfo{year}{n.d.}\natexlab{}.
\newblock \bibinfo{title}{Invgen: Automated Invariant Generation for Smart Contract Fuzzing}.
\newblock \bibinfo{howpublished}{GitHub Repository}.
\newblock
\urldef\tempurl%
\url{https://github.com/fuzzland/invgen}
\showURL{%
\tempurl}
\newblock
\shownote{Accessed: 2025-04-09}.


\bibitem[Grieco et~al\mbox{.}(2020)]%
        {grieco2020echidna}
\bibfield{author}{\bibinfo{person}{Gustavo Grieco}, \bibinfo{person}{Will Song}, \bibinfo{person}{Artur Cygan}, \bibinfo{person}{Josselin Feist}, {and} \bibinfo{person}{Alex Groce}.} \bibinfo{year}{2020}\natexlab{}.
\newblock \showarticletitle{Echidna: effective, usable, and fast fuzzing for smart contracts}.
\newblock \bibinfo{journal}{\emph{Proceedings of the 29th ACM SIGSOFT International Symposium on Software Testing and Analysis}} (\bibinfo{year}{2020}).
\newblock


\bibitem[Group(nd)]%
        {nielsenusability101}
\bibfield{author}{\bibinfo{person}{Nielsen~Norman Group}.} \bibinfo{year}{n.d.}\natexlab{}.
\newblock \bibinfo{booktitle}{\emph{Usability 101: Introduction to Usability}}.
\newblock Nielsen Norman Group.
\newblock
\urldef\tempurl%
\url{https://www.nngroup.com/articles/usability-101-introduction-to-usability/}
\showURL{%
\tempurl}


\bibitem[He et~al\mbox{.}(2021)]%
        {he2021eosafe}
\bibfield{author}{\bibinfo{person}{Ningyu He}, \bibinfo{person}{Ruiyi Zhang}, \bibinfo{person}{Haoyu Wang}, \bibinfo{person}{Lei Wu}, \bibinfo{person}{Xiapu Luo}, \bibinfo{person}{Yao Guo}, \bibinfo{person}{Ting Yu}, {and} \bibinfo{person}{Xuxian Jiang}.} \bibinfo{year}{2021}\natexlab{}.
\newblock \showarticletitle{{EOSAFE}: Security Analysis of {EOSIO} Smart Contracts}. In \bibinfo{booktitle}{\emph{30th {USENIX} Security Symposium ({USENIX} Security 21)}}. \bibinfo{publisher}{USENIX Association}, \bibinfo{pages}{1271--1288}.
\newblock
\urldef\tempurl%
\url{https://www.usenix.org/conference/usenixsecurity21/presentation/he-ningyu}
\showURL{%
\tempurl}


\bibitem[Kelly et~al\mbox{.}(2019)]%
        {kelly2019case}
\bibfield{author}{\bibinfo{person}{Matthew Kelly}, \bibinfo{person}{Christoph Treude}, {and} \bibinfo{person}{Alex Murray}.} \bibinfo{year}{2019}\natexlab{}.
\newblock \showarticletitle{A case study on automated fuzz target generation for large codebases}. In \bibinfo{booktitle}{\emph{Proceedings of the 2019 ACM/IEEE International Symposium on Empirical Software Engineering and Measurement (ESEM'19)}}. \bibinfo{pages}{1--6}.
\newblock


\bibitem[Klooster(2022)]%
        {klooster2022effectiveness}
\bibfield{author}{\bibinfo{person}{et~al. Klooster}.} \bibinfo{year}{2022}\natexlab{}.
\newblock \bibinfo{title}{Effectiveness and Scalability of Fuzzing Techniques in CI/CD Pipelines}.
\newblock
\showeprint[arxiv]{2205.14964}~[cs.SE]


\bibitem[Li et~al\mbox{.}(2018)]%
        {li2018fuzzing}
\bibfield{author}{\bibinfo{person}{Jun Li}, \bibinfo{person}{Bodong Zhao}, {and} \bibinfo{person}{Chao Zhang}.} \bibinfo{year}{2018}\natexlab{}.
\newblock \showarticletitle{Fuzzing: a survey}.
\newblock \bibinfo{journal}{\emph{Cybersecurity}} \bibinfo{volume}{1}, \bibinfo{number}{1} (\bibinfo{year}{2018}), \bibinfo{pages}{1--13}.
\newblock


\bibitem[Li et~al\mbox{.}(2021)]%
        {li2021Unifuzz}
\bibfield{author}{\bibinfo{person}{Yuwei Li}, \bibinfo{person}{Shouling Ji}, \bibinfo{person}{Yuan Chen}, \bibinfo{person}{Sizhuang Liang}, \bibinfo{person}{Wei-Han Lee}, \bibinfo{person}{Yueyao Chen}, \bibinfo{person}{Chenyang Lyu}, \bibinfo{person}{Chunming Wu}, \bibinfo{person}{Raheem Beyah}, \bibinfo{person}{Peng Cheng}, \bibinfo{person}{Kangjie Lu}, {and} \bibinfo{person}{Ting Wang}.} \bibinfo{year}{2021}\natexlab{}.
\newblock \showarticletitle{{UNIFUZZ}: A Holistic and Pragmatic {Metrics-Driven} Platform for Evaluating Fuzzers}. In \bibinfo{booktitle}{\emph{30th USENIX Security Symposium (USENIX Security 21)}}. \bibinfo{publisher}{USENIX Association}, \bibinfo{pages}{2777--2794}.
\newblock
\showISBNx{978-1-939133-24-3}
\urldef\tempurl%
\url{https://www.usenix.org/conference/usenixsecurity21/presentation/li-yuwei}
\showURL{%
\tempurl}


\bibitem[Liang et~al\mbox{.}(2018a)]%
        {liang2018fuzzing}
\bibfield{author}{\bibinfo{person}{Hongliang Liang}, \bibinfo{person}{Xiaoxiao Pei}, \bibinfo{person}{Xiaodong Jia}, \bibinfo{person}{Wuwei Shen}, {and} \bibinfo{person}{Jian Zhang}.} \bibinfo{year}{2018}\natexlab{a}.
\newblock \showarticletitle{Fuzzing: State of the art}.
\newblock \bibinfo{journal}{\emph{IEEE Transactions on Reliability}} \bibinfo{volume}{67}, \bibinfo{number}{3} (\bibinfo{year}{2018}), \bibinfo{pages}{1199--1218}.
\newblock


\bibitem[Liang et~al\mbox{.}(2018b)]%
        {liang2018fuzz}
\bibfield{author}{\bibinfo{person}{Jie Liang}, \bibinfo{person}{Mingzhe Wang}, \bibinfo{person}{Yuanliang Chen}, \bibinfo{person}{Yu Jiang}, {and} \bibinfo{person}{Renwei Zhang}.} \bibinfo{year}{2018}\natexlab{b}.
\newblock \showarticletitle{Fuzz testing in practice: Obstacles and solutions}. In \bibinfo{booktitle}{\emph{Proceedings of the 25th International Conference on Software Analysis, Evolution and Reengineering (SANER'18)}}. \bibinfo{pages}{562--566}.
\newblock


\bibitem[Ma et~al\mbox{.}(2021)]%
        {ma2021pluto}
\bibfield{author}{\bibinfo{person}{Fuchen Ma}, \bibinfo{person}{Zhenyang Xu}, \bibinfo{person}{Meng Ren}, \bibinfo{person}{Zijing Yin}, \bibinfo{person}{Yuanliang Chen}, \bibinfo{person}{Lei Qiao}, \bibinfo{person}{Bin Gu}, \bibinfo{person}{Huizhong Li}, \bibinfo{person}{Yu Jiang}, {and} \bibinfo{person}{Jiaguang Sun}.} \bibinfo{year}{2021}\natexlab{}.
\newblock \showarticletitle{Pluto: Exposing vulnerabilities in inter-contract scenarios}.
\newblock \bibinfo{journal}{\emph{IEEE Transactions on Software Engineering}} \bibinfo{volume}{48}, \bibinfo{number}{11} (\bibinfo{year}{2021}), \bibinfo{pages}{4380--4396}.
\newblock


\bibitem[Melvillian({[n.\,d.]})]%
        {docum1}
\bibfield{author}{\bibinfo{person}{Melvillian}.} \bibinfo{year}{[n.\,d.]}\natexlab{}.
\newblock \bibinfo{title}{Invariant test fuzzed functions always use msg.value == 0}.
\newblock \bibinfo{howpublished}{\url{https://github.com/foundry-rs/foundry/issues/8449}}.
\newblock
\newblock
\shownote{Accessed: 2025-02-15}.


\bibitem[Meta({[n.\,d.]})]%
        {llama}
\bibfield{author}{\bibinfo{person}{Meta}.} \bibinfo{year}{[n.\,d.]}\natexlab{}.
\newblock \bibinfo{title}{LLama 3.3 70B}.
\newblock \bibinfo{howpublished}{\url{https://huggingface.co/meta-llama/Llama-3.3-70B-Instruct/}}.
\newblock
\newblock
\shownote{Accessed: 2025-02-15}.


\bibitem[mistralai({[n.\,d.]})]%
        {mistral}
\bibfield{author}{\bibinfo{person}{mistralai}.} \bibinfo{year}{[n.\,d.]}\natexlab{}.
\newblock \bibinfo{title}{Mixtral-8x7B-Instruct-v0.1}.
\newblock \bibinfo{howpublished}{\url{https://huggingface.co/mistralai/Mixtral-8x7B-Instruct-v0.1}}.
\newblock
\newblock
\shownote{Accessed: 2025-02-15}.


\bibitem[Natella(2022)]%
        {Natella2022}
\bibfield{author}{\bibinfo{person}{Roberto Natella}.} \bibinfo{year}{2022}\natexlab{}.
\newblock \showarticletitle{StateAFL: Greybox fuzzing for stateful network servers}.
\newblock \bibinfo{journal}{\emph{Empirical Software Engineering}} \bibinfo{volume}{27}, \bibinfo{number}{7} (\bibinfo{date}{4 10} \bibinfo{year}{2022}), \bibinfo{pages}{191}.
\newblock
\showISSN{1573-7616}
\href{https://doi.org/10.1007/s10664-022-10233-3}{doi:\nolinkurl{10.1007/s10664-022-10233-3}}


\bibitem[Nourry et~al\mbox{.}(2023)]%
        {nourry2023human}
\bibfield{author}{\bibinfo{person}{Olivier Nourry}, \bibinfo{person}{Yutaro Kashiwa}, \bibinfo{person}{Bin Lin}, \bibinfo{person}{Gabriele Bavota}, \bibinfo{person}{Michele Lanza}, {and} \bibinfo{person}{Yasutaka Kamei}.} \bibinfo{year}{2023}\natexlab{}.
\newblock \showarticletitle{The Human Side of Fuzzing: Challenges Faced by Developers during Fuzzing Activities}.
\newblock \bibinfo{journal}{\emph{ACM Transactions of Software Engineering and Methodology}} (\bibinfo{year}{2023}).
\newblock


\bibitem[Plöger et~al\mbox{.}(2021)]%
        {ploger2021usability}
\bibfield{author}{\bibinfo{person}{Stephan Plöger}, \bibinfo{person}{Mischa Meier}, {and} \bibinfo{person}{Matthew Smith}.} \bibinfo{year}{2021}\natexlab{}.
\newblock \showarticletitle{A Qualitative Usability Evaluation of the Clang Static Analyzer and libFuzzer with CS Students and CTF Players}. In \bibinfo{booktitle}{\emph{Seventeenth Symposium on Usable Privacy and Security (SOUPS'21)}}.
\newblock


\bibitem[Plöger et~al\mbox{.}(2023)]%
        {ploger2023usability}
\bibfield{author}{\bibinfo{person}{Stephan Plöger}, \bibinfo{person}{Mischa Meier}, {and} \bibinfo{person}{Matthew Smith}.} \bibinfo{year}{2023}\natexlab{}.
\newblock \showarticletitle{A Usability Evaluation of AFL and libFuzzer with CS Students}. In \bibinfo{booktitle}{\emph{Proceedings of the 2023 CHI Conference on Human Factors in Computing Systems (CHI '23)}}.
\newblock


\bibitem[Qasse et~al\mbox{.}(2023)]%
        {qasse2023smart}
\bibfield{author}{\bibinfo{person}{Ilham Qasse}, \bibinfo{person}{Mohammad Hamdaqa}, {and} \bibinfo{person}{Bj\"{o}rn~{\TH}\'{o}r J\'{o}nsson}.} \bibinfo{year}{2023}\natexlab{}.
\newblock \showarticletitle{Smart Contract Upgradeability on the Ethereum Blockchain Platform: An Exploratory Study}.
\newblock \bibinfo{journal}{\emph{arXiv preprint arXiv:2304.06568}} (\bibinfo{year}{2023}).
\newblock


\bibitem[Ren et~al\mbox{.}(2021)]%
        {ren2021empirical}
\bibfield{author}{\bibinfo{person}{Meng Ren} {et~al\mbox{.}}} \bibinfo{year}{2021}\natexlab{}.
\newblock \showarticletitle{Empirical evaluation of smart contract testing: what is the best choice?}. In \bibinfo{booktitle}{\emph{Proceedings of the 30th ACM SIGSOFT International Symposium on Software Testing and Analysis (ISSTA 2021)}}.
\newblock


\bibitem[Sayeed et~al\mbox{.}(2020)]%
        {sayeed2020smart}
\bibfield{author}{\bibinfo{person}{Sarwar Sayeed}, \bibinfo{person}{Hector Marco-Gisbert}, {and} \bibinfo{person}{Tom Caira}.} \bibinfo{year}{2020}\natexlab{}.
\newblock \showarticletitle{Smart Contract: Attacks and Protections}.
\newblock \bibinfo{journal}{\emph{IEEE Access}}  \bibinfo{volume}{8} (\bibinfo{year}{2020}), \bibinfo{pages}{24416--24427}.
\newblock


\bibitem[Shou et~al\mbox{.}(2023a)]%
        {shou2023ityfuzz}
\bibfield{author}{\bibinfo{person}{Chaofan Shou}, \bibinfo{person}{Shangyin Tan}, {and} \bibinfo{person}{Koushik Sen}.} \bibinfo{year}{2023}\natexlab{a}.
\newblock \showarticletitle{ItyFuzz: Snapshot-Based Fuzzer for Smart Contract}.
\newblock \bibinfo{journal}{\emph{Proceedings of the 32nd ACM SIGSOFT International Symposium on Software Testing and Analysis}} (\bibinfo{year}{2023}), \bibinfo{pages}{322--333}.
\newblock


\bibitem[Shou et~al\mbox{.}(2023b)]%
        {ityfuzz}
\bibfield{author}{\bibinfo{person}{Chaofan Shou}, \bibinfo{person}{Shangyin Tan}, {and} \bibinfo{person}{Koushik Sen}.} \bibinfo{year}{2023}\natexlab{b}.
\newblock \showarticletitle{ItyFuzz: Snapshot-Based Fuzzer for Smart Contract}. In \bibinfo{booktitle}{\emph{Proceedings of the 32nd ACM SIGSOFT International Symposium on Software Testing and Analysis}} (Seattle, WA, USA) \emph{(\bibinfo{series}{ISSTA 2023})}. \bibinfo{publisher}{Association for Computing Machinery}, \bibinfo{address}{New York, NY, USA}, \bibinfo{pages}{322–333}.
\newblock
\showISBNx{9798400702211}
\href{https://doi.org/10.1145/3597926.3598059}{doi:\nolinkurl{10.1145/3597926.3598059}}


\bibitem[Spencer and Garrett(2009)]%
        {spencer2009card}
\bibfield{author}{\bibinfo{person}{D. Spencer} {and} \bibinfo{person}{J.J. Garrett}.} \bibinfo{year}{2009}\natexlab{}.
\newblock \bibinfo{booktitle}{\emph{Card Sorting: Designing Usable Categories}}.
\newblock \bibinfo{publisher}{Rosenfeld Media}.
\newblock
\showISBNx{9781933820026}
\showLCCN{2009920137}
\urldef\tempurl%
\url{https://books.google.ca/books?id=_h4D9gqi5tsC}
\showURL{%
\tempurl}


\bibitem[unboxedtype({[n.\,d.]})]%
        {docum2}
\bibfield{author}{\bibinfo{person}{unboxedtype}.} \bibinfo{year}{[n.\,d.]}\natexlab{}.
\newblock \bibinfo{title}{Questions regarding functional capabilities of the tool}.
\newblock \bibinfo{howpublished}{\url{https://github.com/crytic/echidna/issues/255}}.
\newblock
\newblock
\shownote{Accessed: 2025-02-15}.


\bibitem[wadealexc({[n.\,d.]})]%
        {config1}
\bibfield{author}{\bibinfo{person}{wadealexc}.} \bibinfo{year}{[n.\,d.]}\natexlab{}.
\newblock \bibinfo{title}{Make it easier to re-run failed fuzz tests with verbose logging}.
\newblock \bibinfo{howpublished}{\url{https://github.com/foundry-rs/foundry/issues/7206}}.
\newblock
\newblock
\shownote{Accessed: 2025-02-15}.


\bibitem[Wu(2024)]%
        {wu2024are}
\bibfield{author}{\bibinfo{person}{et~al. Wu}.} \bibinfo{year}{2024}\natexlab{}.
\newblock \showarticletitle{Are We There Yet? Unraveling the State-of-the-Art Smart Contract Fuzzers}. In \bibinfo{booktitle}{\emph{Proceedings of the IEEE/ACM 46th International Conference on Software Engineering (ICSE '24)}}.
\newblock


\bibitem[Xi et~al\mbox{.}(2024)]%
        {Rui2024PomaBuster}
\bibfield{author}{\bibinfo{person}{Rui Xi}, \bibinfo{person}{Zehua Wang}, {and} \bibinfo{person}{Karthik Pattabiraman}.} \bibinfo{year}{2024}\natexlab{}.
\newblock \showarticletitle{{ POMABuster: Detecting Price Oracle Manipulation Attacks in Decentralized Finance }}. In \bibinfo{booktitle}{\emph{2024 IEEE Symposium on Security and Privacy (SP)}}. \bibinfo{publisher}{IEEE Computer Society}, \bibinfo{address}{Los Alamitos, CA, USA}, \bibinfo{pages}{3923--3942}.
\newblock
\href{https://doi.org/10.1109/SP54263.2024.00257}{doi:\nolinkurl{10.1109/SP54263.2024.00257}}


\bibitem[Yang et~al\mbox{.}(2025)]%
        {yang2025csafuzzer}
\bibfield{author}{\bibinfo{person}{J. Yang}, \bibinfo{person}{X. Zhao}, \bibinfo{person}{H. Zhang}, {et~al\mbox{.}}} \bibinfo{year}{2025}\natexlab{}.
\newblock \showarticletitle{CSAFuzzer: Fuzzing smart contracts combining with static analysis}.
\newblock \bibinfo{journal}{\emph{Empirical Software Engineering}} \bibinfo{volume}{30}, \bibinfo{number}{1} (\bibinfo{year}{2025}), \bibinfo{pages}{62}.
\newblock
\href{https://doi.org/10.1007/s10664-025-10623-3}{doi:\nolinkurl{10.1007/s10664-025-10623-3}}


\bibitem[Ynyesto({[n.\,d.]})]%
        {cross-chain}
\bibfield{author}{\bibinfo{person}{Ynyesto}.} \bibinfo{year}{[n.\,d.]}\natexlab{}.
\newblock \bibinfo{title}{Invariant BSC fork test fails}.
\newblock \bibinfo{howpublished}{\url{https://github.com/foundry-rs/foundry/issues/748}}.
\newblock
\newblock
\shownote{Accessed: 2025-02-15}.


\bibitem[Yun et~al\mbox{.}(2022)]%
        {yun2022fuzzing}
\bibfield{author}{\bibinfo{person}{Joobeom Yun}, \bibinfo{person}{Fayozbek Rustamov}, \bibinfo{person}{Juhwan Kim}, {and} \bibinfo{person}{Youngjoo Shin}.} \bibinfo{year}{2022}\natexlab{}.
\newblock \showarticletitle{Fuzzing of embedded systems: A survey}.
\newblock \bibinfo{journal}{\emph{Comput. Surveys}} (\bibinfo{year}{2022}).
\newblock


\end{thebibliography}
